\documentclass[aip,amsmath,amssymb,reprint,]{revtex4-1}
\usepackage{amsmath}
\usepackage{amssymb}
\usepackage{bbold}
\usepackage{etoolbox}
\usepackage{breqn}
\usepackage{graphicx}
\usepackage{subfigure}
\usepackage{dcolumn}
\usepackage{bm}
\usepackage{float}
\usepackage{hyperref}
\usepackage{xcolor}

\makeatletter
\patchcmd\eq@setnumber{\stepcounter}{\refstepcounter}{}{%
  \errmessage{Patching \noexpand\eq@setnumber failed}%
}
\makeatother
\makeatletter
\let\cat@comma@active\@empty
\makeatother
\let \oldnabla \nabla
\renewcommand{\nabla}{\bm{\oldnabla}}
\renewcommand{\vec}{\mathbf}

\newcommand{\vrho}{\bm \rho}

\newcommand{\om}{\omega}
\newcommand{\id}{\mathbb{1}}

\newcommand{\vx}{\bm{\sigma}}

\newcommand{\vecx}{\vec{x}}

\newcommand{\vc}{\vec{c}}
\newcommand{\mC}{\vec{C}}
\newcommand{\vv}{\vec{v}}

\newcommand{\scM}{\mathcal{M}}

\newcommand{\Kcm}{K_c^{(-)}}
\newcommand{\Kcu}{K_c^{(u)}}
\newcommand{\Kcp}{K_c^{(+)}}

\newcommand{\pholder}{\mathord{\color{black!33}\bullet}}%
\newcommand{\uu}[1]{\mathbf{#1}}
\newcommand{\wt}[1]{\widetilde{#1}}
\DeclareMathOperator{\Tr}{Trace}
\DeclareMathOperator{\diag}{diag}
\DeclareMathOperator{\newRe}{Re}
\DeclareMathOperator{\newIm}{Im}

\renewcommand{\Re}{\newRe}
\renewcommand{\Im}{\newIm}

\begin{document}

\title{Observing Microscopic Transitions from Macroscopic Bursts: Instability-Mediated Resetting in the Incoherent Regime of the $D$-dimensional Generalized Kuramoto Model}
\author{Sarthak Chandra}
\affiliation{Institute for Research in Electronics and Applied Physics, University of Maryland, College Park, Maryland 20742, U.S.A.}
\author{Edward Ott}
\affiliation{Institute for Research in Electronics and Applied Physics, University of Maryland, College Park, Maryland 20742, U.S.A.}

\begin{abstract}
This paper considers a recently introduced $D$-dimensional generalized Kuramoto model for many $(N\gg 1)$ interacting agents in which the agents states are $D$-dimensional unit vectors. It was previously shown that, for even (but not odd) $D$, similar to the original Kuramoto model ($D=2$), there exists a continuous dynamical phase transition from incoherence to coherence of the time asymptotic attracting state (time $t\to\infty$) as the coupling parameter $K$ increases through a critical value which we denote $K_c^{(+)}>0$. We consider this transition from the point of view of the stability of an incoherent state, where an incoherent state is defined as one for which the $N\to\infty$ distribution function is time-independent and the macroscopic order parameter is zero. In contrast with $D=2$, for even $D>2$ there is an infinity of possible incoherent equilibria, each of which becomes unstable with increasing $K$ at a different point $K=K_c$. Although there are incoherent equilibria for which $K_c=K_c^{(+)}$, there are also incoherent equilibria with a range of possible $K_c$ values below $K_c^{(+)}$, $(K_c^{(+)}/2)\leq K_c < K_c^{(+)}$. How can the possible instability of incoherent states arising at $K=K_c<K_c^{(+)}$ be reconciled with the previous finding that, at large time $(t\to \infty)$, the state is always incoherent unless $K>K_c^{(+)}$? We find, for a given incoherent equilibrium, that, if $K$ is rapidly increased from $K<K_c$ to $K_c<K<K_c^{(+)}$, due to the instability, a short, macroscopic burst of coherence is observed, in which the coherence initially grows exponentially, but then reaches a maximum, past which it decays back into incoherence. Furthermore, after this decay, we observe that the equilibrium has been reset to a new equilibrium whose $K_c$ value exceeds that of the increased $K$. Thus this process, which we call `Instability-Mediated Resetting,' leads to an increase in the effective $K_c$ with continuously increasing $K$, until the equilibrium has been effectively set to one for which for which $K_c\approx K_c^{(+)}$. Thus Instability-Mediated Resetting leads to a unique critical point of the $t\to\infty$ time asymptotic state ($K=K_c^{(+)}$) in spite of the existence of an infinity of possible pretransition incoherent states.
\end{abstract}

\maketitle
\begin{quotation}
The dynamical phase transition from incoherence to coherence for a recently proposed, higher-dimensional generalization of the Kuramoto model, is examined from the point of view of the stability of the incoherent state. It is found that, due to the higher dimensionality, there is a continuum of different possible pretransition incoherent equilibrium states, each with distinct stability properties. This, in turn, leads to a novel phenomenon, which we call `Instability-Mediated Resetting,' which enables the existence of a unique critical transition point in spite of the infinite continuum of possible pretransition states. 
In general, these results provide an example illustrating that, for systems with a large number of entities described via a macroscopic variable(s), a degeneracy of microscopic states corresponding to the same macroscopic variable may occur, and that signatures of such a degeneracy may be observable in the transient macroscopic system dynamics.
\end{quotation}

\section{Introduction}\label{sec:I}
\subsection{Background}\label{sec:Ia}
Motivated by a host of applications, much recent research has been focused on efforts aimed at understanding the behavior of large systems of many interacting dynamical agents. An important tool elucidating issues in this general area has been the study of simplified paradigmatic models. A prime example of such a model is the Kuramoto model\cite{Kuramoto1975,Acebron2005,Strogatz2000,Ott2002},
\begin{equation}\label{eq:2dkuramoto}
d\theta_i/dt = \om_i + \frac{K}{N} \sum_{j=1}^N \sin(\theta_j - \theta_i),
\end{equation} 
where $N$ is the number of agents ($i=1,2,\hdots,N$), $\theta_i$ is an angle variable that specifies the state of agent $i$, the parameter $K$ characterizes the coupling strength, and $\omega_i$ is the natural frequency of agent $i$ ($\dot\theta_i=\om_i$ in the absence of coupling), where $\om_i$ is typically chosen randomly for each $i$ from some distribution function $g(\om)$ ($\int g(\om)d\om = 1$).
Because the parameter $\omega_i$ characterizing the dynamics of each agent $i$ is different for each agent, the agents are said to be heterogeneous. This model and its many generalizations have been used to study a wide variety of applications and phenomena. Examples include synchronously flashing fireflies\cite{Buck1976}, cellular clocks in the brain\cite{Lu2016}, Josephson junction circuits\cite{Wiesenfeld1998}, pedestrian-induced oscillation of foot bridges\cite{Eckhardt2007}, and motion direction alignment in large groups of agents (e.g., drones or flocking animals)\cite{Moshtagh2007,Zhu2013a,Wang2005}, among many others. In the first four of these examples $\theta_i$ represents the phase angle of an oscillation experienced by agent $i$, while, in contrast, in the fifth example, $\theta_i$ specifies the direction in which agent $i$ moves.

One aspect of the Kuramoto model and is previous generalizations is that the state of agent $i$ is given by the single scalar angle variable $\theta_i(t)$. Recently, a generalization of these models has been introduced in which the state of the agent $i$ is a $D$-dimensional unit vector, $\vx_i(t)$, thus allowing for more degrees of freedom in the dynamics of the individual agents. In this generalized model the $D$-dimensional unit vector, $\vx_i(t)$, is taken to evolve according to the real equation\cite{Chandra2018,Olfati-Saber2006,Zhu2013},
\begin{equation}\label{eq:Ddimkuramoto}
d\vx_i/dt = K [\vrho - (\vrho\cdot\vx_i)\vx_i] + \uu{W}_i\vx_i,
\end{equation}
where the $D$-dimensional vector $\vrho(t)$ (to be specified subsequently) is a common field felt by all the agents, and $\uu{W}_i$ (analogous to $\om_i$ in Eq. (\ref{eq:2dkuramoto})) is a $D\times D$ antisymmetric matrix ($\uu{W}_i^T = -\uu{W}_i$) which we refer to as the rotation rate matrix. Note that for $K=0$ Eq. (\ref{eq:2dkuramoto}) becomes $\dot\vx_i = \uu{W}_i \vx_i$ which represents a uniform rate of rotation of $\vx_i$ in $D$-dimensional space, $\vx_i(t) = [\exp(\uu{W}_i t)]\vx_i(0)$, analogous to the action of the frequency $\om_i$ in $D=2$. Dotting Eq. (\ref{eq:Ddimkuramoto}) with $\vx_i$, we obtain $d|\vx_i|^2/dt=0$, as required by our designation of $\vx_i$ as a unit vector. In general, depending on the situation to be modeled, $\vrho(t)$ can be chosen in different ways\cite{Ott2008,Chandra2018}. In this paper we focus on the simplest interesting choice,
\begin{equation}\label{eq:ordpar}
\vrho(t) = \frac{1}{N}\sum_{i=1}^N \vx_i(t),
\end{equation}
and we call $|\vrho(t)|$, the `order parameter.' We note that, as shown in Ref.\onlinecite{Chandra2018}, Eqs. (\ref{eq:Ddimkuramoto}) and (\ref{eq:ordpar}) reduce to Eq. (\ref{eq:2dkuramoto}) for $D=2$ with
\begin{equation*}
\vx_i = \begin{pmatrix}\cos\theta_i \\ \sin\theta_i\end{pmatrix} \text{ and } \uu{W}_i = \begin{pmatrix} 0 & -\om_i \\ \om_i & 0\end{pmatrix},
\end{equation*}
thus justifying Eqs. (\ref{eq:Ddimkuramoto}) and (\ref{eq:ordpar}) as a `generalization' of the Kuramoto model, Eq. (\ref{eq:2dkuramoto}), to higher dimensionality. One motivation for this generalization is the previously mentioned example of the application of Eq. (\ref{eq:2dkuramoto}) to model motion alignment in flocks: For $D=2$ (equivalent to the standard Kuramoto case, Eq. (\ref{eq:2dkuramoto})) the direction of agent motion (characterized by the scalar angle $\theta_i$ or the unit vector $(\cos\theta_i \; \sin\theta_i)^T$) can be described for agents moving along a two-dimensional surface (like the surface of the Earth), while, if the agents are, e.g., moving in three dimensions (as for drones flying in the air), then the direction of an agent's motion ($\vx_i$ for agent $i$) is necessarily given by a three-dimensional unit vector. In addition, Ref. \onlinecite{Olfati-Saber2006} has considered the dynamics of the vector $\vx_i$ as characterizing the evolution of the opinions of an individual within a group of interacting individuals as the group evolves towards consensus. Another interesting point\cite{Chandra2018} is that the inter-agent coupling for Eqs. (\ref{eq:Ddimkuramoto}) and (\ref{eq:ordpar}) is the same as that for the classical, mean-field, zero-temperature, Heisenberg model for the evolution of $N$ interacting spin states $\vx_i$ in the presence of frozen-in random site disorder (the terms $\uu{W}_i\vx_i$, with $\uu{W}_i$ randomly chosen).

Based on our previous work (see Ref. \onlinecite{Chandra2018}) we view Eqs. (\ref{eq:Ddimkuramoto}) and (\ref{eq:ordpar}) as the simplest $D$-dimensional generalization of the Kuramoto model subject to the assumption that the state of any agent is a unit vector. See Sec. IV of Ref. \onlinecite{Chandra2018} for a generalization, motivated by flocking drones, in which the agents are regarded as $D$-dimensional extended-body agents whose states of orientation are described by $(D-1)$ mutually perpendicular unit vectors. Although the model in Sec. IV of Ref. \onlinecite{Chandra2018} is quite different from that considered here, Ref. \onlinecite{Chandra2018} shows that it shares the same qualitative type of transition behavior as Eq. (\ref{eq:Ddimkuramoto}). Thus we conjecture that the model we study in the present paper can provide a general guide to the possible behavior of other related systems.

\subsection{The Rotation Rates $\uu{W}_i$}\label{sec:Ib}

Equation (\ref{eq:Ddimkuramoto}) with zero rotation rate ($\uu{W}_i=0$) or a uniform rotation rate ($\uu{W}_i=\uu{W}$) was introduced in Refs. \onlinecite{Olfati-Saber2006, Zhu2013}. The generalization to heterogeneous rotation rates\cite{Chandra2018} (the situation to be considered in the present paper) makes Eq. (\ref{eq:Ddimkuramoto}) more similar to the original Kuramoto model and widens its range of applicability. In what follows, as in Ref.\onlinecite{Chandra2018}, we assume that the rotation rate matrix $\uu{W}_i$ is randomly generated for each $i$, by choosing each of its $D(D-1)/2$ upper triangular matrix elements, $w_{pq}^{(i)}$ (with $p<q$), independently from a zero-mean, Gaussian distribution function as described in Sec. \ref{sec:2}.
Alternately, we can say that each of the $\uu{W}_i$ is randomly drawn from the ensemble of random antisymmetric matrices corresponding to the Gaussian distribution. It is important to note that this ensemble is invariant under rotations; i.e., the ensemble is unchanged when every matrix in the ensemble is subjected to the same rotation, $\uu{W}\to\uu{R}\uu{W}$, for any orthogonal matrix $\uu{R}$ (e.g., Ref.\onlinecite{Mehta1968}). 

\subsection{The $N\to\infty$ limit and the multiplicity of incoherent equilibria}\label{sec:Ic}

We are interested in the case where $N\gg 1$, and, to facilitate analysis, we consider the limit $N\to\infty$, for which we characterize the system state for dimensionality $D$ by a distribution function $F(\uu{W},\vx,t)$ such that the fraction of the agents lying in the differential volume element $d\vx d\uu{W}$ centered at $(\vx,\uu{W})$ in $\vx$-$\uu{W}$ space is $F(\uu{W},\vx,t) d\vx d\uu{W}$ at time $t$. Throughout this paper we will use the term \emph{state} to denote this system state characterized by $F(\uu{W},\vx,t)$. We define an incoherent equilibrium state to be such that $\partial F/\partial t=0$ and $|\vrho|=0$, where, since we consider the limit $N\to\infty$, Eq. (\ref{eq:ordpar}) is replaced by
\begin{equation}\label{eq:distordpar}
\vrho(t) = \int\int F(\uu{W},\vx,t) \vx d\vx d\uu{W}.
\end{equation}
As shown in Sec.\ref{sec:2}, for $D>2$ there is an infinite continuum of equilibrium (i.e., time-independent) distribution functions $F$ for which $|\vrho|=0$.  We can think of these distributions as defining a manifold $\scM$ in the space of distribution functions.

Within this manifold, a given $F$ is neutrally stable to a perturbation $\delta F$ such that $F+\delta F$ also lies in $\scM$. Section \ref{sec:3} is devoted to an analysis of the stability of the manifold $\scM$; i.e., what happens if $\delta F$, the perturbation to $F$, is transverse to $\scM$. Before discussing what we find in Sec. \ref{sec:3} for the case $D>2$, it is first useful to recall the well-known results for the original Kuramoto model, corresponding to $D=2$, as well as relevant results from Ref.\onlinecite{Chandra2018} for $D>2$.

\subsection{The Dynamical phase transition}\label{sec:Id}

\begin{figure}
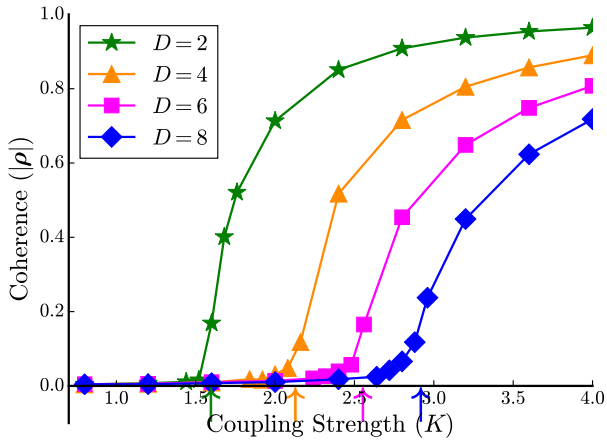

\includegraphics[width=\columnwidth]{{{Phase_transitions_even_dim_IMR}}}
\caption{Dynamical Phase Transition for the Generalized Kuramoto model for $D=2$ (green stars), $4$ (orange triangles), $6$ (magenta squares) and $8$	 (blue diamonds) dimensions. The $|\vrho|$ values indicated by the plotted markers are obtained by choosing the values of $\vx_i(t=0)$ and $\uu{W}_i$ for each of the $N=10^5$ agents randomly (where the probability distribution of $\vx_i(0)$ is isotropic in direction and that of $\uu{W}_i$ is as given in Sec. \ref{sec:2}) and then integrating Eq. (\ref{eq:Ddimkuramoto}) from each such initial condition until $|\vrho(t)|$ attains a steady value. These steady state values attained appeared to be independent of this choice of initial condition.
The theoretical predictions from Ref. \onlinecite{Chandra2018} for the critical coupling strength, $K_c^{(+)}$, above which stable $|\vrho|>0$ steady states exist are indicated by correspondingly colored vertical arrows on the $x$-axis. 
}
\label{fig:rhovK}
\end{figure} 

In the case of the original ($D=2$) Kuramoto model, Eq. (\ref{eq:2dkuramoto}), for $N\to\infty$ one can consider a distribution function in $(\om,\theta)$; i.e., $F(\uu{W},\vx,t)\to f(\om,\theta,t)$. In this $D=2$ case, in contrast to the $D>2$ generalized model, Eq. (\ref{eq:Ddimkuramoto}), there is only one $|\vrho|=0$ equilibrium distribution function, namely $f=g(\om)/(2\pi)$. Furthermore, it has long been well-established for $D=2$, that, as $K$ increases continuously from zero, the long-time ($t\to\infty$) stable value of the order parameter $|\vrho|$ undergoes a continuous transition from incoherence ($|\vrho|=0$) to partial coherence ($0<|\vrho|<1$) as $K$ passes a critical value that depends on $g(\om)$, see the green curve marked by the star symbols in Fig. \ref{fig:rhovK}. We denote this critical value by $K_c^{(+)}$. This transition has been studied from two different points of view (see Refs.\onlinecite{Kuramoto1975,Acebron2005,Strogatz2000,Ott2002}):

Method (i): It is assumed that $f$ reaches a steady state ($\partial f/\partial t=0$) and the resulting nonlinear equation for $f$ is then analytically solved, yielding two possible solutions for the order parameter $|\vrho|$; one has $|\vrho|=0$ and corresponds to $f=g(\om)/(2\pi)$; the other satisfies a transcendental equation for $|\vrho|$ as a function of $K$ involving an integral of the $\om$-distribution function $g$. Taking $g$ to be continuous, unimodal, symmetric, and peaked at $\om=0$, the transcendental root for $|\vrho|$ only exists for $K\geq K_c^{(+)}>0$ and gives the $|\vrho|>0$ branch in Fig. \ref{fig:rhovK}. In this approach, an analytical result for $K_c^{(+)}$ is obtained by taking the limit $|\vrho|\to0^+$ in the expression for the transcendental branch. This is the approach originally taken by Kuramoto, who then essentially assumed that the $|\vrho|=0$ branch applies for $K\leq K_c^{(+)}$, and the $|\vrho|>0$ branch applies for $K>K_c^{(+)}$.

Method (ii): Considering the $|\vrho|=0$ equilibrium state, a linear stability analysis was applied\cite{Kuramoto1975,Acebron2005,Strogatz2000,Ott2002,Strogatz1991}, and it was found that the $|\vrho|=0$ equilibrium state (which exists for \emph{all} $K$) becomes unstable when $K$ increases through a critical value which is the same as that found for $K_c^{(+)}$ by method (i).

Thus the value of $K_c^{(+)}$ for the original Kuramoto problem can be obtained straightforwardly by following either method (i) or method (ii).

In Ref.\onlinecite{Chandra2018} using Method (i), previously employed for the original Kuramoto problem, analysis giving the critical transition values for even $D$ were obtained. These values are indicated by the vertical arrows in Fig. \ref{fig:rhovK}, and agree well with the plotted numerical results. 

Parenthetically, we note that for odd $D\geq 3$, which is \emph{not} considered in this paper, the transition is qualitatively different from that shown in Fig. \ref{fig:rhovK}. Namely, as shown in Ref. \onlinecite{Chandra2018}, when $D$ is odd, as $K$ increases from negative values through zero there is a discontinuous jump in the coherence $|\vrho|$.

\subsection{Linear Stability of the incoherent state}\label{sec:Ie}

Motivated by the results in Fig. \ref{fig:rhovK}, in Sec. \ref{sec:3} we report results of a stability analysis of the incoherent equilibria for even $D$ greater than two. That is, we attempt an analysis similar to method (ii), previously applied to the original Kuramoto model. We find that the straightforward correspondence that applies for $D=2$ between the method (i) result for $K_c^{(+)}$ and the method (ii) stability result does not hold for $D=4,6,8\hdots$, and that the apparent paradox presented by this finding is resolved by a novel phenomenon that we call \emph{Instability-Mediated Resetting} (IMR).

Specifically, our stability analysis in Sec. \ref{sec:3} applied to the infinity of possible incoherent equilibrium states found in Sec. \ref{sec:2}, shows that different incoherent equilibria have different stability properties. Considering one such incoherent equilibrium, as $K$ increases, the equilibrium will become unstable as $K$ passes through some value $K_c$ which depends on the specific incoherent equilibrium considered. There are thus many possible values of $K_c$, in fact we find a continuum of such $K_c$ values spanning a range between $(K_c^{(+)}/2)$ and $K_c^{(+)}$.

\subsection{Instability-Mediated Resetting (IMR)}\label{sec:If}

These stability results for $D=4,6,\hdots$ suggest the following question. How can instability of incoherent equilibrium states for $K<K_c^{(+)}$ be reconciled with the numerical results of Fig. \ref{fig:rhovK} and the corresponding method (i) analytical results (the vertical arrows in Fig. \ref{fig:rhovK})? The answer to this question is given in Sec. \ref{sec:4} which reports the following results on the nonlinear evolution of the instability found in Sec. \ref{sec:3}: Considering an incoherent equilibrium which becomes unstable at $K=K_c<K_c^{(+)}$, if one starts with $K<K_c$ and then rapidly increases $K$ to lie in the range $K_c<K<K_c^{(+)}$, the order parameter $|\vrho|$ initially experiences growth consistent with the existence of instability. This growth, however, slows as $|\vrho|$ reaches a maximum, and subsequently decays back to zero. But, after this short-lived macroscopic burst, once $|\vrho|$ returns to essentially zero the resulting incoherent equilibrium is different from that which existed before the instability occurred, and this resulting new incoherent equilibrium loses stability only at a value of the coupling strength between the value that $K$ has been increased to and $K_c^{(+)}$.

In fact, if the initial burst occurred due to a value of $K$ roughly in the middle of the range $K_c<K<K_c^{(+)}$, the resulting equilibrium may be one which loses stability only at $K_c^{(+)}$ itself, i.e., upon further increase of $K$, $|\vrho|$ remains near zero until $K$ increases past $K_c^{(+)}$. If $K$ is suddenly increased through $K_c^{(+)}$ there is unstable growth of $|\vrho|$, as for when $K$ is increased suddenly through $K_c$, but now $|\vrho|$ asymptotically approaches a positive value consistent with Fig. \ref{fig:rhovK} for $K>K_c^{(+)}$. The essential point is that the instability for $K_c<K<K_c^{(+)}$ resets the equilibrium to a new state which is stable for $K<K_c^{(+)}$ and becomes unstable only when $K$ exceeds $K_c^{(+)}$, consistent with the plot (Fig. \ref{fig:rhovK}) of the $t\to\infty$ order parameter vs $K$. This is the IMR phenomenon previously referred to.

\subsection{Main points of this paper}\label{sec:Ih}

This paper focuses on the case of even dimensional generalizations of the Kuramoto model of the form Eq. (\ref{eq:Ddimkuramoto}). A main message of this paper is that, although the curves, $|\vrho(t\to\infty)|$ versus $K$ plotted in Fig. \ref{fig:rhovK} for $D=4,6,\hdots$, are qualitatively similar to the curve for $D=2$, the transient dynamics of $\vrho(t)$ starting from a given incoherent distribution at $t=0$ are surprisingly different for even $D\geq 4$ as compared with $D=2$.	
We will demonstrate in Sec. \ref{sec:2} that for even $D>2$, in contrast to $D=2$, there is an infinite continuum of incoherent stable equilibria in the limit of $N\to\infty$. In Sec. \ref{sec:3} we will perform a linear stability analysis of these equilibria, and show that these equilibria have different critical coupling strengths, i.e., values of $K$ beyond which the equilibria are unstable. Further, we also show that in a continuous range of $K$, each value of $K$ corresponds to the critical coupling strength of some incoherent equilibrium. The upper limit of this range corresponds to earlier results for the critical coupling strength for the $t\to\infty$ macroscopic phase transition of the order parameter shown in Fig. \ref{fig:rhovK}. To reconcile these lower values of critical stability coupling strengths for incoherent equilibria, with the phase transition of Fig. \ref{fig:rhovK}, we will examine the dynamics of the incoherent equilibria beyond their critical coupling strengths. This examination results in the observation of short-lived macroscopic bursts of $|\vrho|$ which lead to the phenomenon of Instability-Mediated Resetting, which we demonstrate and describe in Sec. \ref{sec:4}. We also discuss the effect of finite $N$ on the evolution of these incoherent equilibria in Sec. \ref{sec:4}. 


\section{Incoherent Equilibria}\label{sec:2}

We reiterate that in this paper we will only consider the case of even $D$. For each $\uu{W}$ there are $D/2$ two-dimensional invariant subspaces for the $|\vrho|=0$ evolution equation 
\begin{equation}\label{eq:rho0eqn}
d\vx/dt = \uu{W}\vx.
\end{equation}
To see this, we define the rotation $\uu{R}_D$ to be a $D\times D$ orthogonal matrix that puts $\uu{W}$ in block-diagonal form,
\begin{equation}\label{eq:Wtilde}
\uu{R}_D^T \uu{W} \uu{R}_D = \wt{\uu{W}} = %
\begin{pmatrix}
0 			& \om_1 &   		&		 	  &  &  &  \\
-\om_1 	& 0 		&   		&   		&  &  &  \\
 			 	&   		& 0 		& \om_2 &  &  &  \\
 			 	&   		&-\om_2	& 0     &  &  &  \\
 			 	&   		&     	&       & \ddots &  &  \\
 			 	&   		&     	&       & 			 & 0 & \om_{D/2} \\
 			 	&   		&     	&       & 			 & -\om_{D/2} &0 \\
 
\end{pmatrix},
\end{equation}
with $\om_k$ real. Furthermore, we define $\uu{P}_k$ to be the projection operator that projects a $D$-vector onto the $k^{\text{th}}$ invariant subspace of $\uu{W}$, i.e., $\uu{R}_D^T \uu{P}_k \uu{R}_D = \wt{\uu{P}_k}$ has all elements zero except for the $(2k-1)^{\text{th}}$ and $(2k)^{\text{th}}$ elements on the diagonal which are set to $1$. By construction
\begin{equation*}
\mathbb{1} = \sum_{k=1}^{D/2} \wt{\uu{P}_k} = \sum_{k=1}^{D/2} \uu{P}_k,
\end{equation*}
where $\mathbb{1}$ is the $D$-dimensional identity matrix. Setting $\wt{\vx}=\uu{R}_D^T\vx$ transforms the $|\vrho|=0$ evolution equation Eq. (\ref{eq:rho0eqn}) to
\begin{equation}\label{eq:tildex}
d\wt{\vx}/dt = \wt{\uu{W}}\wt{\vx}.
\end{equation}
Thus, for each $k$, 
\begin{equation}\label{eq:Ck}
C_k = (\wt{\uu{P}_k}\wt{\vx})^T(\wt{\uu{P}_k}\wt{\vx}) = \vx^T \uu{P}_k^T \uu{P}_k \vx = \vx^T \uu{P}_k \vx
\end{equation}
is a constant of motion for the $|\vrho|=0$ evolution equation $d\vx/dt = \uu{W}\vx$. 

Since we are interested in the case where the number of agents, $N$, is large, $N\gg 1$, it is appropriate to simplify the analysis by considering the limit $N\to\infty$, in which case the state of the system can be described by a distribution function, $F(\uu{W},\vx,t)$ satisfying
\begin{equation}\label{eq:Fevolution}
\partial F/\partial t + \nabla_S\cdot (\vv F) =0, \quad \vv=K[\mathbb{1}_D - \vx \vx^T]\vrho + \uu{W}\vx,
\end{equation}
where $\nabla_S\cdot(\vv F)$ represents the divergence of the vector field $\vv F$ on the spherical surface $|\vx|=1$.
Hence any time independent distribution function, $\overline{F_0}(\uu{W},\vx)$, for the $|\vrho|=0$ dynamics must satisfy
\begin{equation}\label{eq:F0bar}
\nabla_S\cdot [(\uu{W}\vx)\overline{F_0}] = (\uu{W}\vx)\cdot \nabla_S \overline{F_0} = 0,
\end{equation}
where $\nabla_S$ represents the gradient operator on the spherical surface $|\vx|=1$. The first equality follows from the fact that $\nabla_S\cdot(\uu{W}\vx) =0$ for $\uu{W}$ an antisymmetric matrix. Since
\begin{equation}\label{eq:Ckconstant}
0=dC_k/dt = \nabla_S C_k(\vx) \cdot d\vx/dt = (\uu{W}\vx)\cdot\nabla_S C_k(\vx),
\end{equation}
by comparing Eqs. (\ref{eq:F0bar}) and (\ref{eq:Ckconstant}), we see that the most general solution for a time-independent distribution $\overline{F_0}$ is
\begin{equation}
\overline{F_0}(\uu{W},\vx) = F_0(\uu{W};C_1,\hdots,C_{D/2})=F_0(\uu{W},\vc),
\end{equation}
where $\vc$ denotes the $(D/2)$-vector $(C_1,\hdots,C_{D/2})^T$, i.e., $\overline{F_0}$ depends on $\uu{W}$ and the $(D/2)$ constants of the motion. There are two constraints. The first one is that, since $|\vx|=1$, we have that $|\vc|=1$. The second constraint is that
\begin{equation}\label{eq:rho0constraint}
0=\vrho_0 = \int \int_{|\vx|=1} \vx \overline{F_0}(\uu{W},\vx) d\uu{W} d\vx,
\end{equation}
which is automatically satisfied if, as we henceforth assume, $\overline{F_0}$ is isotropic in the sense that 
\begin{equation}
\overline{F_0}(\uu{R}^T\uu{W}\uu{R},\uu{R}\vx) = \overline{F_0}(\uu{W},\vx)
\end{equation}
for any rotation matrix $\uu{R}$. Thus
\begin{equation}\label{eq:F0isotropy}
F_0(\uu{R}^T\uu{W}\uu{R},\vc) = F_0(\uu{W},\vc),
\end{equation}
since the constants $C_k$ are invariant to such rotations. Equation (\ref{eq:rho0constraint}) for $\vrho_0$ then yields $\vrho_0=\uu{R}\vrho_0$ for any rotation $\uu{R}$, which then implies that the integral $\int \vx \overline{F_0} d\uu{W}d\vx =0$, as required by our definition of an incoherent state, Eq. (\ref{eq:rho0constraint}).

In our work we consider the case where the marginal distribution of $\uu{W}$ expressed in terms of the matrix elements
\begin{equation}
w_{ii}=0,\quad w_{ij}=-w_{ji}
\end{equation}
is Gaussian. That is,
\begin{dmath}
G(\uu{W})d\uu{W} = \left[ \int_{|\vx|=1} F_0(\uu{W},\vc) d\vx \right] d\uu{W} = \prod_{j=1}^D \prod_{i>j}^D g_M(w_{ij}) dw_{ij},
\end{dmath}
where $g_M(w)$ is the Gaussian distribution
\begin{equation}
g_M(w) = \frac{1}{\sqrt{2\pi \langle w^2 \rangle}} e^{-\frac{w^2}{2 \langle w^2 \rangle}},\quad\langle w^2 \rangle =\int_{-\infty}^{\infty} w^2 g_M(w)dw.
\end{equation}
Thus
\begin{dmath}\label{eq:Gdist}
G(\uu{W}) = (2\pi \langle w^2 \rangle )^{-D(D-1)/4}\exp\left[-\Tr(\uu{W}^T\uu{W})/(4\langle w^2 \rangle) \right].
\end{dmath}
Since $\Tr(\uu{W}^T\uu{W}) = -\Tr(\uu{W}^2)$ is invariant to rotations of $\uu{W}$ (i.e., $\uu{W}\to\uu{R}^T\uu{W}\uu{R}$) and $d\uu{W}=d(\uu{R}\uu{W})$ (since $\det(\uu{R})$=1), we see that $G(\uu{W})$ as defined above is isotropic in the sense that
\begin{equation}\label{eq:Gisotropy}
G(\uu{W})=G(\uu{R}^T\uu{W}\uu{R})
\end{equation}
for any $D\times D$ rotation matrix $\uu{R}$.

According to random matrix theory, the distribution of block frequencies $\om_k$ in Eq. (\ref{eq:Wtilde}) for such a Gaussian ensemble of even-dimensional random antisymmetric matrices with $\langle w^2\rangle$ set to 1 is\cite{Mehta1968}
\begin{dmath}\label{eq:gtilde}
\wt{g}(\om_1,\hdots,\om_{D/2}) \\= \kappa\prod_{1\leq j\leq k \leq D/2} (\om_j^2 - \om_k^2)^2\exp\left(-\sum_{i=1}^{D/2}\om_i^2/2\right),
\end{dmath}
where $\kappa$ is a constant chosen to ensure that the integral of the distribution $\wt{g}(\om_1,\hdots,\om_{D/2})$ is normalized to 1. Note that $\wt{g}$ is symmetric to interchanges of any two of its arguments.

As an aside, we also mention that using Eq. (\ref{eq:Wtilde}), $\uu{W}=\uu{R}_D\wt{\uu{W}}\uu{R}_D^T$, an alternative representation of $G(\uu{W})d\uu{W}$ is
\begin{equation*}
\wt{g}(\om_1,\hdots,\om_{D/2})d\om_1\hdots d\om_{D/2}d\mu(\uu{R}_D),
\end{equation*}
where $\mu$ is the Haar measure for $D\times D$ rotation matrices. (The Haar measure for rotation matrices essentially gives a formal rigorous specification of what we loosely refer to as isotropy\cite{Faraut2008}. In what follows we use our informal notion of `isotropy' and do not invoke Haar measures.)

Returning to the distribution function $F_0$, we define $\hat{F}_0$ by
\begin{equation}
F_0(\uu{W},\vc)=G(\uu{W})\hat{F}_0(\uu{W},\vc),
\end{equation}
where 
\begin{equation}
\int \hat{F}_0 (\uu{W},\vc) \delta(|\vx|-1)d\vx =1.
\end{equation}
Note that $|\vx|^2 = C_1+\hdots+C_{D/2}=1$. Clearly, even with $G(\uu{W})$ specified as Gaussian, there is still an infinity of choices for $\hat{F}_0$ and hence $F_0$. These choices specify how $\vx$ is distributed over the $D/2$ subspaces of $\uu{W}$ that are invariant for the $|\vrho|=0$ dynamics of $\vx$.

\section{Stability of incoherent equilibria}\label{sec:3}

We linearize Eq. (\ref{eq:Ddimkuramoto}) about states corresponding to incoherent equilibria, i.e., $|\vrho|=0$, by setting $\vx=\vx_0+\delta \vx$ and $\vrho=\delta\vrho$ for small perturbations $\delta \vx$ and $\delta\vrho$. This yields,
\begin{align}
d\vx_0/dt &= \uu{W}\vx_0, \label{eq:fp} \\
d \delta\vx/dt &= K[\id - \vx_0 \vx_0^T]\delta\vrho + \uu{W}\delta\vx. \label{eq:perturbation}
\end{align}
Transforming Eq. (\ref{eq:fp}) to the basis that block-diagonalizes $\uu{W}$ (as in Eq. (\ref{eq:Wtilde})), we obtain
\begin{equation}
d\wt{\vx}_0/dt = \wt{\uu{W}} \wt{\vx}_0.
\end{equation}
Thus each two-dimensional subspace $k$ will undergo independent rotation with frequencies corresponding to real $\om_k$ frequencies of $\wt{\uu{W}}$. This gives the solution
\begin{equation}\label{eq:wtvxevolution}
\wt{\vx}_0(t) = \uu{Q}(t) \wt{\vx}_0(0),
\end{equation}
where $\uu{Q}(t)$ is a block diagonal matrix with $(D/2)$ blocks of dimensions $2\times 2$ given by
\begin{equation}\label{eq:Qt}
\uu{Q}(t) = %
\begin{pmatrix}
\uu{Q}_1(t) & & & & \\
      & \ddots & & & \\
			&   & \uu{Q}_k(t) & & \\
			& & & \ddots & \\
			& & & & \uu{Q}_{D/2}(t)
\end{pmatrix},
\end{equation}
with
\begin{equation}\label{eq:Qkt}
\uu{Q}_k(t) = %
\begin{pmatrix}
\cos\om_k t	&\sin\om_k t\\
-\sin\om_k t&\cos\om_k t
\end{pmatrix}
\end{equation}
for $1\leq k \leq D/2$. We can equivalently represent Eq. (\ref{eq:wtvxevolution}) as
\begin{equation}\label{eq:vecxevolution}
\vecx_k(t) = \uu{Q}_k(t) \vecx_k(0)
\end{equation}
for each $k$, where $\vecx_k(t)$ is the two-dimensional vector formed by the $(2k-1)$ and $2k$ components of $\wt{\vx}_0$.

Now, assuming that $\delta\vrho(t) = e^{st} \delta\vrho(0)$, Eq. (\ref{eq:perturbation}) yields
\begin{dmath}\label{eq:dxdrho}
\delta\vx(t) = K \left\{ \int_{-\infty}^t e^{\uu{W}(t-\tau)}  \left[ \id - \vx_0(\tau)\vx_0(\tau)^T\right]e^{s\tau} d\tau \right\} \delta\vrho(0),
\end{dmath}
where $\vx_0(\tau)=\uu{R}_D \uu{Q}(t) \uu{R}_D^T \vx_0(0)$.

We note that the order parameter of the perturbed system, $\delta\vrho$, will be given by the average of $\delta\vx$ over each agent (corresponding to an average over all $\uu{W}$). We also perform an ensemble average over all choices of initial conditions corresponding to a given incoherent equilibrium characterized by $\hat{F}_0(\uu{W},\vc)$. Thus
\begin{equation}\label{eq:drhodx}
\delta\vrho(t) = \langle \langle \delta\vx(t)\rangle_{\vx_0(0)}\rangle_{\uu{W}},
\end{equation}
where $\langle \pholder \rangle_{\vx_0(0)}$ denotes an average over $\vx_0(0)$ at fixed $\uu{W}$, and $\langle \pholder \rangle_{\uu{W}}$ denotes an average over $\uu{W}$. We first average Eq. (\ref{eq:dxdrho}) over $\vx_0(0)$:
\begin{dmath}\label{eq:dxdrhovx0}
\langle \delta\vx(t) \rangle_{\vx_0(0)}= K \left\{ \int_{-\infty}^t e^{\uu{W}(t-\tau)}  \left[ \id - \langle \vx_0(\tau)\vx_0(\tau)^T \rangle_{\vx_0(0)}\right]e^{s\tau} d\tau \right\} \delta\vrho(0).
\end{dmath}
We focus on the evaluation of the term 
\begin{align}
\langle & \vx_0(\tau)\vx_0(\tau)^T \rangle_{\vx_0(0)} \\
                                &= \uu{R}_D\langle \wt{\vx}_0(\tau)\wt{\vx}_0(\tau)^T \rangle_{\vx_0(0)} \uu{R}_D^T, \label{eq:xxtaverage0}\\
                                &= \uu{R}_D \left[ \int_{|\vx|=1} \hat{F}_0(\uu{W},\vc) \wt{\vx}_0(\tau)\wt{\vx}_0(\tau)^T d\vx \right]\uu{R}_D^T. \label{eq:xxtaverage}
\end{align}
Note that $\vx_0 \vx_0^T$ is a $D\times D$ matrix which can be constructed from $(D/2)\times(D/2)$ blocks of $2\times 2$ matrices, where the block at index $(k,l)$ will be $\vecx_k \vecx_l^T$ for $1\leq k,l \leq D/2$.
Defining $\vecx_k(0)=(y_k^+, y_k^-)^T$, we obtain from Eq. (\ref{eq:wtvxevolution})
\begin{equation}
\vecx_k(\tau) = \begin{pmatrix}  y_k^+ \cos\om_k \tau + y_k^-\sin\om_k t \\ -y_k^+\sin\om_k t + y_k^-\cos\om_k t  \end{pmatrix}.
\end{equation}
Since $C_k = (y_k^+)^2 + (y_k^-)^2$, we write
\begin{equation}
y_k^+ = \sqrt{C_k} \cos\theta_k, \quad y_k^- = \sqrt{C_k} \sin\theta_k.
\end{equation}
Thus 
\begin{equation}
\vecx_k(\tau) = \sqrt{C_k} \begin{pmatrix}  \cos(\om_k \tau-\theta_k)\\ \sin(\om_k \tau-\theta_k) \end{pmatrix}.
\end{equation}
We interpret the average to be performed in Eq. (\ref{eq:xxtaverage}) as an average over $\theta_k$ and $\sqrt{C_k}$ for each $k$, with the differential element $d\vx$ transforming to $\prod_k \sqrt{C_k}d\sqrt{C_k} d\theta_k$. 

Noting that $\langle \vecx_k \rangle$ averaged over $\theta_k$ is zero, we see that $\langle \vecx_k \vecx_l^T \rangle$ can only be nonzero if $k=l$. Further, in averaging $\vecx_k \vecx_k^T$, the diagonal terms corresponding to $C_k \cos^2(\om_k \tau-\theta_k)$ and $C_k \sin^2(\om_k \tau-\theta_k)$ will yield $(C_k/2)$ when averaged over $\theta_k$, and the cross terms corresponding to $C_k \sin(\om_k \tau-\theta_k) \cos(\om_k \tau-\theta_k)$ will yield zero. Thus, we obtain
\begin{equation}
\langle \vecx_k \vecx_k^T \rangle_{\theta_k} = \frac{C_k}{2} \id_2,
\end{equation}
where $\id_2$ represents the $2\times 2$ identity matrix. Note that the average over $\theta_k$ removes all $\tau$ dependence in Eq. (\ref{eq:xxtaverage}). Performing the average over $C_k$, we obtain
\begin{equation}
\langle \vecx_k \vecx_k^T \rangle_{\vx_0(0)} = \frac{\bar{C}_k(\uu{W})}{2} \id_2,
\end{equation}
where
\begin{equation}\label{eq:Cbar}
\bar{C}_k(\uu{W}) = \frac{\int_{\Gamma} \hat{F}_0(\uu{W},\vc) C_k d\vc}{\int_{\Gamma} \hat{F}_0(\uu{W},\vc) d\vc},
\end{equation}
with the domain $\Gamma$ corresponding to the set of all $\vc$ such that $0\leq C_k \leq 1$ for all $k$, and $\sum_k C_k=1$.

Thus the quantity $\langle  \vx_0(\tau)\vx_0(\tau)^T \rangle_{\vx_0(0)}$ in Eq. (\ref{eq:xxtaverage0}) becomes
\begin{equation}\label{eq:xxtvx0}
\langle  \vx_0(\tau)\vx_0(\tau)^T \rangle_{\vx_0(0)} = \uu{R}_D \bar{\mC}(\uu{W}) \uu{R}_D^T,
\end{equation}
where $\bar{\mC}(\uu{W})$ is the $D$-dimensional diagonal matrix,
\begin{dmath*}
\bar{\mC}(\uu{W}) = \frac{1}{2}\diag\left[ \bar{C}_1(\uu{W}),\bar{C}_1(\uu{W}),\bar{C}_2(\uu{W}),\bar{C}_2(\uu{W}),\\ \hdots,\bar{C}_{D/2}(\uu{W}),\bar{C}_{D/2}(\uu{W}) \right].
\end{dmath*}

Now performing the average over $\uu{W}$ as prescribed in Eq. (\ref{eq:drhodx}), we obtain from Eqs. (\ref{eq:dxdrhovx0}) and (\ref{eq:xxtvx0})
\begin{dmath*}
\delta\vrho(t)=\delta\vrho(0)e^{st}= K \left\{ \int d\uu{W} G(\uu{W})\int_{-\infty}^t e^{\uu{W}(t-\tau)}  \uu{R}_D\left[ \id - \bar{\mC}\right]\uu{R}_D^Te^{s\tau} d\tau \right\} \delta\vrho(0),
\end{dmath*}
or
\begin{dmath*}
\left\{\id - K \int d\uu{W}G(\uu{W}) \int_{-\infty}^{t} e^{(t-\tau)(\uu{W}-s\id)} \uu{R}_D \left[\id - \bar{\mC}(\uu{W}) \right]\uu{R}_D^T d\tau \right\} \delta\vrho(0)=0.
\end{dmath*}
Integrating over $\tau$, we obtain
\begin{dmath*}
\left\{\id - K \int d\uu{W}G(\uu{W})  (s\id-\uu{W})^{-1} \uu{R}_D \left[\id - \bar{\mC}(\uu{W}) \right]\uu{R}_D^T  \right\} \delta\vrho(0)=0.
\end{dmath*}
Using the change of basis Eq. (\ref{eq:Wtilde}),
\begin{dmath}\label{eq:intermediate}
\left\{\id - K \int d\uu{W}G(\uu{W})  \uu{R}_D (s\id-\wt{\uu{W}})^{-1}  \left[\id - \bar{\mC}(\uu{W}) \right]\uu{R}_D^T  \right\} \delta\vrho(0)=0.
\end{dmath}
Since $\uu{R}_D(\uu{W})=\uu{R}_D(-\uu{W})$, $G(\uu{W})=G(-\uu{W})$, and by Eq. (\ref{eq:F0isotropy}) $\bar{\mC}(\uu{W})=\bar{\mC}(-\uu{W})$, we can replace the $(s\id-\wt{\uu{W}})^{-1}$ term in Eq. (\ref{eq:intermediate}) by
\begin{equation}\label{eq:simplifying}
\frac{1}{2}\left[\frac{1}{s\id-\wt{\uu{W}}} + \frac{1}{s\id+\wt{\uu{W}}}\right] = \frac{s}{s^2\id - \wt{\uu{W}}^2}.
\end{equation}
Noting that
\begin{equation*}
\begin{pmatrix} 0 & -\om \\ \om & 0 \end{pmatrix}^2 = -\om^2 \id_2,
\end{equation*}
the quantity in Eq. (\ref{eq:simplifying}) becomes
\begin{dmath}
H(s;\om_1,\hdots,\om_{D/2}) \\= s \diag\left[ \frac{1}{s^2 + \om_1^2},\frac{1}{s^2 + \om_1^2},\hdots,\frac{1}{s^2 + \om_{D/2}^2},\frac{1}{s^2 + \om_{D/2}^2}\right],
\end{dmath}
which when inserted into Eq. (\ref{eq:intermediate}), yields
\begin{dmath}\label{eq:RVRdeltarho}
\left\{\id - K \int d\uu{W}G(\uu{W})  \uu{R}_D \uu{V} \uu{R}_D^T  \right\} \delta\vrho(0)=0,
\end{dmath}
where
\begin{align*}
\uu{V} &= H(s;\om_1,\hdots,\om_{D/2}) \left[\id - \bar{\mC}(\uu{W}) \right] \\
       &= s \diag\left[  \frac{1-\frac{\bar{C}_1}{2}}{s^2 + \om_1^2},\frac{1-\frac{\bar{C}_1}{2}}{s^2 + \om_1^2},\right. \\ 
			&\left. \hdots,\frac{1-\frac{\bar{C}_{D/2}}{2}}{s^2 + \om_{D/2}^2},\frac{1-\frac{\bar{C}_{D/2}}{2}}{s^2 + \om_{D/2}^2}\right].
\end{align*}

Noting that $G(\uu{W})$ is isotropic in the sense of Eq. (\ref{eq:Gisotropy}), we can average $\uu{R}_D \uu{V} \uu{R}_D^T$ (equivalently $\uu{V}$) over an isotropic ensemble of rotations and replace $d\uu{W}G(\uu{W})$ by the distribution of the rotation invariant quantities characterizing $\uu{W}$, i.e., $\{\om_1,\hdots,\om_{D/2}\}$. Noting that $\Tr(\uu{V})=\Tr(\uu{R} \uu{V} \uu{R}^T)$ for any rotation $\uu{R}$ and that the average $\langle \uu{R} \uu{V} \uu{R}^T \rangle_{\uu{R}}$ over an isotropic ensemble of rotations $\uu{R}$ must, by the isotropy, be a scalar multiple of the $D\times D$ identity matrix, we obtain
\begin{align}
\langle \uu{R} \uu{V} \uu{R}^T \rangle_{\uu{R}} &= \frac{1}{D}\Tr(\uu{V})\id \nonumber \\
                                                &= \left(\frac{2s}{D}\sum_{k=1}^{D/2} \frac{1-\bar{C}_k(\uu{W})/2}{s^2+\om_k^2}\right)\id. \label{eq:RVRT}
\end{align}
Using Eqs.(\ref{eq:RVRT}) and (\ref{eq:gtilde}), we find that, for $\delta\vrho(0)\neq 0$, Eq. (\ref{eq:RVRdeltarho}) yields the scalar equation
\begin{dmath}\label{eq:scalardispersion}
1-\frac{2Ks}{D} \int d\om_1\hdots\int d\om_{D/2} \wt{g}(\om_1,\hdots,\om_{D/2}) \\ \times\sum_{k=1}^{D/2} \frac{1-\bar{C}_k(\uu{W})/2}{s^2+\om_k^2} = 0,
\end{dmath}
where after averaging over the ensemble of rotations, we have replaced $G(\uu{W})d\uu{W}$ in Eq. (\ref{eq:RVRdeltarho}) by 
\begin{equation*}
\wt{g}(\om_1,\hdots,\om_{D/2})d\om_1\hdots d\om_{D/2},
\end{equation*}
with $\wt{g}$ being the distribution of block frequencies (Eq.(\ref{eq:gtilde})) corresponding to the distribution $G(\uu{W})$. Note that, by the invariance of $\hat{F}_0(\uu{W},\vc)$ with respect to rotations of $\uu{W}$, although in our definition of $\bar{C}_k$ we write $\bar{C}_k\equiv\bar{C}_k(\uu{W})$ (see Eq. (\ref{eq:Cbar})), we can more specifically write it as a function only of the rotation invariant block frequencies $\{\om_1,\hdots,\om_{D/2}\}$ characterizing $\uu{W}$:
\begin{equation*}
\bar{C}_k(\uu{W})\to\bar{C}_k(\om_1,\hdots,\om_{D/2}).
\end{equation*}
Due to the isotropy of the ensemble of matrices $\uu{W}$, the function $\bar{C}_k(\om_1,\hdots,\om_{D/2})$ will be invariant to any swapping of indices, i.e.,
\begin{equation*}
\bar{C}_k(\om_1,\hdots,\om_k,\hdots,\om_{D/2}) = \bar{C}_1(\om_k,\hdots,\om_1,\hdots,\om_{D/2}),
\end{equation*}
for all $k$. Since $\wt{g}$ is also invariant to swapping of its arguments (see Eq. (\ref{eq:gtilde})), we obtain
\begin{dmath}\label{eq:scalardispersionfinal}
1-Ks \int d\om_1\hdots\int d\om_{D/2} \wt{g}(\om_1,\hdots,\om_{D/2}) \\ \times \frac{1-\bar{C}_1(\om_1,\hdots,\om_k,\hdots,\om_{D/2})/2}{s^2+\om_1^2} = 0,
\end{dmath}

To obtain $K_c$, the critical coupling constant at instability onset, we consider the limit $\Re(s)\to 0$ from $\Re(s)>0$. Denoting the real and imaginary parts of $s$ by $p$ and $q$ respectively, we hence consider the limit of $s=p+iq\to iq$ from $p>0$. Note that
\begin{align}
\lim_{p\to0^+} \frac{s}{s^2 + \om_1^2} &= \lim_{p\to0^+} \frac{1}{2}\left\{\frac{-i}{\om_1-i(p+iq)} + \frac{i}{\om_1+i(p+iq)} \right\} \nonumber \\
                                        &= \pi[\delta(\om_1+q)+\delta(\om_1-q)]/2,
\end{align}
where $\delta(x)$ represents the Dirac delta function at $x$. Thus we find from Eq. (\ref{eq:scalardispersion}) that
\begin{dmath}
1-\frac{K_c(q) \pi}{2} \\ \times\int \wt{g}(\om_1,\hdots,\om_{D/2}) \left[1-\frac{\bar{C}_1(\om_1,\hdots,\om_{D/2})}{2}\right] \\ \times[\delta(\om_1+q)+\delta(\om_1-q)] \prod_{j=1}^{D/2}d\om_j=0,
\end{dmath}
where $K_c(q)$ is the critical coupling strength at which a small perturbation to the distribution $F_0$ begins to have an unstable mode growing as $e^{st}$ with $\Im(s)=q$. 
Given that our choice of an isotropic ensemble of rotation matrices $\uu{W}$, the functions $\wt{g}$ and $\bar{C}_1$ must be even functions in each of their arguments.
Thus, 
\begin{dmath}\label{eq:Kcfinal}
K_c(q) = \frac{1}{\pi}\left[ \int \wt{g}(q,\om_2,\hdots\om_{D/2}) \\ \times \left[1-\frac{1}{2}\bar{C}_1(q,\om_2,\hdots\om_{D/2})\right]\prod_{j\geq 2}d\om_j \right]^{-1}.
\end{dmath}
The $q$ dependence of $K_c$ indicates that for each value of $q$ there exists a mode of instability that arises at the corresponding value of $K_c(q)$. However, the critical coupling strength $K_c$ of a distribution $F_0$ is the smallest value of $K$ for which there is an unstable mode. Thus
\begin{equation}
K_c=\min_{q} K_c(q).
\end{equation}
For notational convenience we define 
\begin{equation}\label{eq:hom}
h(\om) = \int d\om_2 \hdots \int d\om_{D/2} \wt{g}(\om,\om_2,\hdots,\om_{D/2}).
\end{equation}
Recalling Eq. (\ref{eq:Cbar}), we see that $\bar{C}_1$ is the expected value of the fraction of $|\vx_i|^2$ lying in the first invariant subspace of $\uu{W}$. Hence, for $D\geq 4$,
\begin{equation*}
0 \leq \bar{C}_1(\om_1,\hdots,\om_{D/2}) \leq 1
\end{equation*}
for all realizations of $\uu{W}$. For the case of $D=2$ (i.e., the standard Kuramoto model) there is only a single frequency associated with $\uu{W}$, and hence $\bar{C}_1=1$. Thus, Eq. (\ref{eq:Kcfinal}) shows that $K_c(q)$ must lie  in the range
\begin{equation}\label{eq:Kcqbound}
\frac{1}{\pi h(q)} \leq K_c(q) \leq \frac{2}{\pi h(q)}.
\end{equation}
Following the form of $\wt{g}$ given in Eq. (\ref{eq:gtilde}), we observe that $h(q)$ is maximized $q=0$. Thus, minimizing each of the three terms in the above inequality,
\begin{equation}\label{eq:Kcrange}
0 \leq K_c^{(-)}=\frac{1}{\pi h(0)} \leq K_c \leq K_c^{(+)}=\frac{2}{\pi h(0)}.
\end{equation}
Using the above inequality we make the following observations:
\begin{itemize}
	\item For all incoherent equilibria, the corresponding $K_c$ is greater than $K_c^{(-)}$. Thus any incoherent equilibrium will be stable for coupling strengths $K<K_c^{(-)}$
	\item There does not exist any incoherent equilibrium distribution whose $K_c$ is greater than $K_c^{(+)}$. Thus, all incoherent equilibria become unstable for coupling strengths $K>K_c^{(+)}$. This is consistent with Fig. \ref{fig:rhovK}, where we see that for $K>K_c^{(+)}$ the system attains an equilibria with $|\vrho|>0$.
	\item For an arbitrary choice of $\bar{C}_k$ it is not necessary that $K_c(q)$ will be minimized at $q=0$. However, for several of the examples we consider below we will consider simple choices for $\bar{C}_1$ such that the minima will occur at $K_c(0)$.
	\item The inequality in Eq. (\ref{eq:Kcrange}) does not have an explicit $D$ dependence. However, as noted above for $D=2$, $\bar{C}_1=1$, resulting in a single critical coupling constant $K_c = K_c^{(+)} = 2/(\pi h(0))=2/(\pi \wt{g}(0))$.
\end{itemize}

In the subsequent discussion we will consider the special case of $D=4$ and give examples of distributions and their corresponding critical coupling strengths for the onset of instability.

\emph{Uniform $\vx$}: For each $\uu{W}_i$, the corresponding unit vector $\vx_i$ is chosen randomly with uniform probability in all directions. Thus the expected value of the magnitude squared of the projection $\vx_i\uu{P}_k\vx_i$ onto subspace $k$ (see Eq. (\ref{eq:Ck})) is the same for all of the $D/2$ subspaces, and, since $|\vx_i|^2=1$, this expected value is $(2/D)$, i.e., 
\begin{equation}\label{eq:Cu}
\bar{C}_1=2/D. 
\end{equation} 
The uniform distribution is of particular interest because of its ease of implementation in computer simulations and because of the intuitive naturalness of this choice. From Eq. (\ref{eq:Kcfinal}) we obtain 
\begin{align*}
K_c(q) &= \frac{1}{\pi}\left[ \int \wt{g}(q,\om_2,\hdots\om_{D/2})\frac{(D-1)}{D} \prod_{j\geq 2}d\om_j \right]^{-1}, \\
       &= \frac{D}{(D-1)\pi h(q)},
\end{align*}
and hence
\begin{equation}
K_c^{(u)}=\frac{D}{(D-1)\pi h(0)},
\end{equation}
giving $K_c^{(u)}=4/[3\pi h(0)]$ for $D=4$.

\emph{Minimally Stable Distribution}: We define a minimally stable distribution to be one whose critical coupling constant for the onset of instability corresponds to the lower bound of Eq. (\ref{eq:Kcrange}), i.e., $K_c=K_c^{(-)}$. 
To construct such a distribution we initialize each agent arbitrarily but restricted to the subspace that is orthogonal to the subspace corresponding to the smallest absolute value of the frequency, i.e., for each agent we set $C_{min}=0$ where 
\begin{equation}
C_{min}=C_k \text{ if }|\om_k|\leq |\om_j|\text{ for all }1\leq j \leq (D/2).
\end{equation} 
For $D=4$ this corresponds to 
\begin{equation}\label{eq:Cminus}
\bar{C}_1(\om_1,\om_2)= \begin{cases}0 	&\;\text{if }|\om_1|\leq|\om_2|, \\ 1  &\;\text{if }|\om_1|>|\om_2|. \end{cases}
\end{equation}
Note that for this distribution $\bar{C}_1(0,\om_2)=0$ for all $\om_2$. To see why this results in a minimally stable distribution we compute the integral in Eq. (\ref{eq:Kcfinal}) and observe that $K_c(q)$ for this distribution is minimized at $q=0$ (see Fig. \ref{fig:Kcq}; For this minimally stable distribution $K_c(q)$ has been labelled as $K_c^{(min)}(q)$, shown in purple). This gives $K_c=K_c(0)=1/(\pi h(0))=K_c^{(-)}$.

\emph{Maximally Stable Distribution}: We define a maximally stable distribution to be one whose critical coupling constant for the onset of instability corresponds to the upper bound of Eq. (\ref{eq:Kcrange}), i.e., $K_c=K_c^{(+)}$.
In $D=4$, such a distribution can be set up similar to the case of the minimally stable distribution, by choosing the $\vx_i$ to lie entirely in the subspace corresponding to the smallest absolute value of the frequency, i.e., by setting $C_{min}=1$ for each agent\footnote{An analogous construction of setting $C_{min}=1$ for each agent does not work to construct a maximally stable distribution in $D\geq 6$. While this implies that $C_{min}=1$ for each agent is not always a maximally stable distribution, it does \emph{not} imply that there is no such distribution in $D\geq 6$. We leave the construction of such a distribution to future work.}. This corresponds to
\begin{equation}\label{eq:Cplus}
\bar{C}_1(\om_1,\om_2)= \begin{cases}1 	&\;\text{if }|\om_1|\leq|\om_2|, \\ 0  &\;\text{if }|\om_1|>|\om_2|. \end{cases}
\end{equation}
As earlier, integration of Eq. (\ref{eq:Kcfinal}) with the above $\bar{C}_1$ results in an expression for $K_c(q)$ which is again minimized at $q=0$ (see Fig. \label{fig:Kcq}; For this maximally stable distribution $K_c(q)$ has been labelled as $K_c^{(max)}(q)$, shown in green). This gives $K_c=K_c(0)=2/(\pi h(0))=K_c^{(+)}$.

In addition to yielding an upper bound on $K_c$, maximally stable distributions are of particular interest because they surprisingly tend to arise naturally in our numerical simulations performed on necessarily finite system size, even when other equilibrium distributions $F_0(\uu{W},\vx)$ are initialized (e.g., when the uniform $\vx$ distribution is initialized); see Sec. \ref{sec:4finite}. Note that it is not necessary for a maximally stable distribution to have $C_{min}=1$ for each agent; for example, the maximally stable distributions attained due to the long-time limit of finite-$N$-effects as shown in Fig. \ref{fig:finiteN} do not have $C_{min}=1$ for each agent.

\begin{figure}
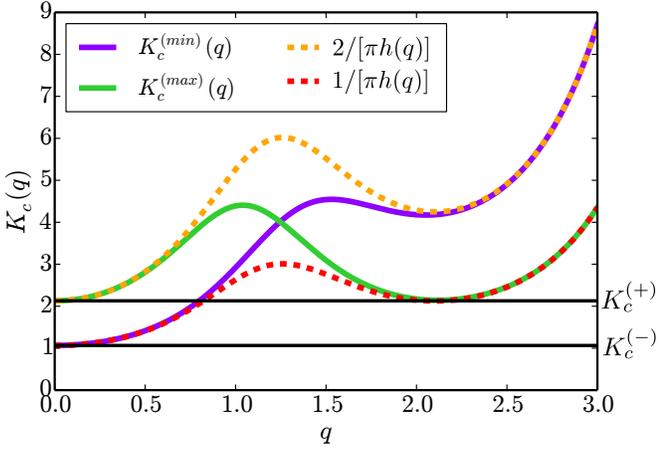

\includegraphics[width=\columnwidth]{{{Kcq_test}}}
\caption{$K_c(q)$ vs $q$ for the case of the minimally stable distribution (shown in green, labelled $K_c^{(min)}(q)$) corresponding to Eq. (\ref{eq:Cminus}) and the maximally stable distribution (shown in purple, labelled $K_c^{(max)}(q)$) corresponding to Eq. (\ref{eq:Cplus}) for $D=4$. Note that $K_c(q)$ is always bounded by $1/[\pi h(q)]$ (shown as the red dashed curve) and $2/[\pi h(q)]$ (shown as the orange dashed curve) as indicated in Eq. (\ref{eq:Kcqbound}). The critical coupling strength for the onset of instability, $K_c$ for a given distribution is given by the minimum value attained by $K_c(q)$, which for $K_c^{(min)}(q)$ and $K_c^{(max)}(q)$ is at $q=0$. ($K_c^{(max)}(q)$ appears to be approximately minimized at $q\approx 2.12$, corresponding to a value of $K_c(q)=2.141$. The true minima however is at $q=0$, corresponding to $K_c(q)=2.128$)
}
\label{fig:Kcq}
\end{figure} 

The largest possible value of the critical coupling constant, $K_c^{(+)}$, beyond which no stable incoherent equilibria exist, corresponds to the calculation of $K_c$ performed in Ref. \onlinecite{Chandra2018} for $D\geq 4$, as shown via the arrows marked in Fig.\ref{fig:rhovK}.
Thus, for $D\geq 4$ we obtain 
\begin{equation}\label{eq:Kcsmup}
K_c^{(-)}=\frac{1}{\pi h(0)} < K_c^{(u)}=\frac{D}{(D-1)\pi h(0)} < K_c^{(+)}=\frac{2}{\pi h(0)}
\end{equation}

In particular, for the choice of the distribution of rotations matrices Eq. (\ref{eq:Gdist}), Ref. \onlinecite{Chandra2018}, presents an expression for $h(0)$, which we use to give values of $\Kcm$, $\Kcu$ and $\Kcp$ for the cases of even $D\leq 8$ in Table \ref{tab:Kcs}.

\begin{table}%
\begin{tabular}{c||c|c|c|c}
       & $h(0)$                     & $\Kcm$  & $\Kcu$   & $\Kcp$        \\ \hline
$D=2$  & $(2\pi)^{-1/2}$              & N/A     & 1.596    & 1.596         \\
$D=4$  & $(3/4)\times(2\pi)^{-1/2}$   & 1.064   & 1.418    & 2.128         \\
$D=6$  & $(5/8)\times(2\pi)^{-1/2}$   & 1.277   & 1.532    & 2.553         \\
$D=8$  & $(35/64)\times(2\pi)^{-1/2}$ & 1.459   & 1.667    & 2.918         \\
\end{tabular}
\caption{Expressions for $h(0)$ and numerical values of $\Kcm$, $\Kcu$ and $\Kcp$ for $D=2$, $4$, $6$ and $8$. The expression for $h(0)$ is obtained from Ref.\onlinecite{Chandra2018}, and the values of the various critical coupling strengths are obtained from Eq. (\ref{eq:Kcsmup})}
\label{tab:Kcs}
\end{table}

In order to demonstrate that any $K_c$ value between $K_c^{(-)}$ and $K_c^{(+)}$ can occur depending on the equilibrium, we consider a particular simple example: For every $\uu{W}_i$
 in our randomly chosen $\uu{W}$-ensemble, we determine $\vx_i$ according to either one of the three protocols specified above with probabilities $p^{(u)}$ (for the uniform case), $p^{(+)}$ (for the maximally stable case) or $p^{(-)}$ (for the minimally stable case), with 
\begin{equation} \label{eq:pconstraint}
p^{(u)} + p^{(-)} + p^{(+)}=1
\end{equation}
Using the expected value interpretation of $\bar{C}_1$, we thus obtain from Eqs. (\ref{eq:Cu}), (\ref{eq:Cminus}) and (\ref{eq:Cplus})
\begin{equation*}
\bar{C}_1 = \begin{cases}p^{(+)}+2p^{(u)}/D 	&\;\text{if }|\om_1|\leq|\om_2|, \\ p^{(-)}+2p^{(u)}/D  &\;\text{if }|\om_1|>|\om_2|, \end{cases}
\end{equation*}
corresponding to 
\begin{equation}\label{eq:Kc3ps}
K_c = p^{(u)} K_c^{(u)} + p^{(+)} K_c^{(+)} + p^{(-)} K_c^{(-)}.
\end{equation}
Hence for $D=4$, by choosing values of $p^{(u,+,-)}$, we can construct a distribution to have any given value of $K_c$ between $K_c^{(-)}$ and $K_c^{(+)}$ (We expect a similar construction to exist for all even $D\geq 4$). Furthermore, for any given $K_c$ in the range Eq. (\ref{eq:Kcrange}), Eqs. (\ref{eq:pconstraint}) and (\ref{eq:Kc3ps}) represent only two constraints on the three parameters, $p^{(u)}$, $p^{(-)}$ and $p^{(+)}$. Thus, for each value of $K_c$ in the range $K_c^{(-)} < K_c < K_c^{(+)}$ there are an infinity of possible choices for $p^{(u)}$, $p^{(+)}$ and $p^{(-)}$ (i.e., an infinite number of distribution functions) satisfying Eq. (\ref{eq:Kc3ps}).

\section{Macroscopic bursts and Instability-Mediated Resetting}\label{sec:4}

In the previous section we considered $N\to\infty$ and showed that, for even $D\geq 4$, the stability of incoherent equilibria depends on their associated equilibrium distribution function, $F$. 
In particular, we observe a range of critical parameter values $K_c$ for instability onset from $K_c^{(-)}$ to $\Kcp = 2\Kcm$, where $K_c$ depends on $F$. By definition, for any $K<\Kcm$ all incoherent states are stable, and for $K>\Kcp$ all incoherent states are unstable. 

We now return to the central question posed in Sec. \ref{sec:I}, namely, how can we reconcile the loss of the stability of an incoherent state at a critical coupling value of $K_c<\Kcp$ with Fig. \ref{fig:rhovK}, which shows the attracting value of the magnitude of the order parameter, $|\vrho|$, characterizing the coherence of the agent population for $t\to\infty$? In particular, in Fig. \ref{fig:rhovK}, how is the time-asymptotic value for $|\vrho|$ maintained at zero for $\Kcm<K<\Kcp$ despite multiple incoherent equilibria losing their stabilities at critical coupling strengths $K_c<K$?

\begin{figure*}
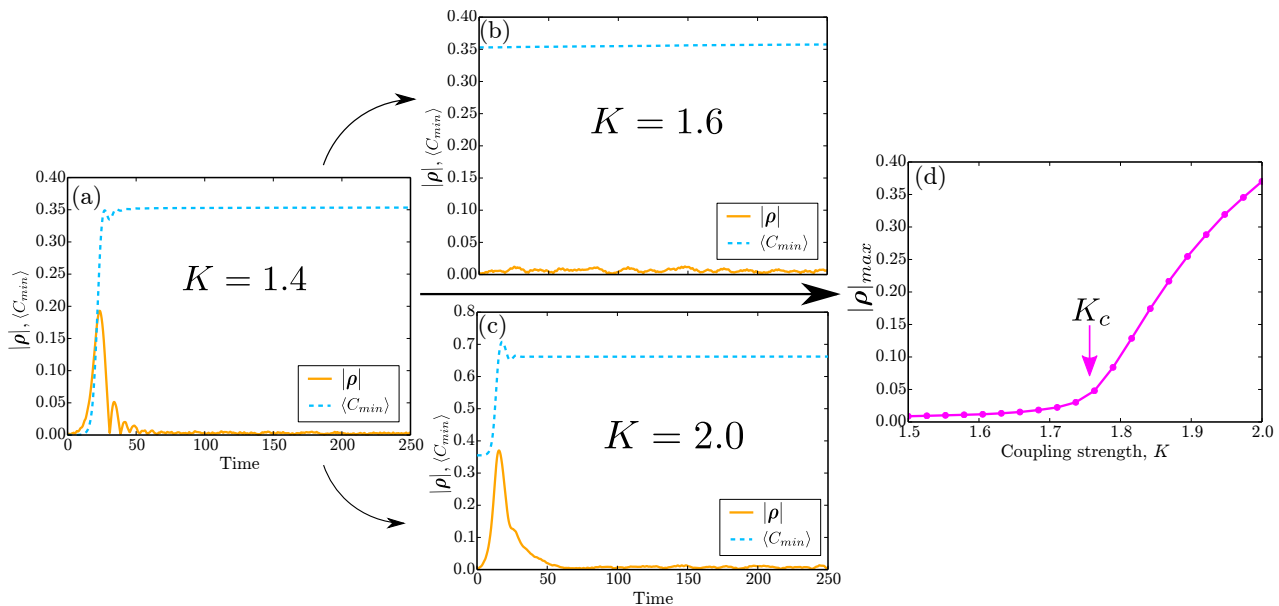

\includegraphics[width=2.\columnwidth]{{{Bump_K1.4_combined}}}
\caption{Representative plots demonstrating the short-lived macroscopic burst of coherence and the resulting IMR. (a) The magnitude of the order parameter (orange solid curve) and $\langle C_{min}\rangle$ (blue dashed curve) as a function of time for a system setup with a minimally stable distribution corresponding to $\langle C_{min} \rangle=0$ and evolved with $K=1.4$. Note the sharp rise and fall of $|\vrho|$, i.e. the macroscopic burst of coherence, accompanied by the increase of the value of $\langle C_{min} \rangle$ (i.e., IMR). This results in an increase of the critical coupling constant for instability onset of the new incoherent state. Panels (b) and (c) show the order parameter evolution beginning with the distribution function at the last time-step of (a) but with $K$ increased to $K=1.6$ and $2.0$ respectively. The presence of a macroscopic burst of $|\vrho|$ in (c) and not in (b) indicates that $K_c$ has been reset to a value between $1.6$ and $2.0$. In panel (d) $|\vrho|_{max}$ indicates the largest value of $|\vrho|$ for systems initialized similar to (b) or (c) following a discontinuous increase of the coupling constant to a value $K$ plotted on the horizontal axis. $|\vrho|_{max}$ is macroscopically observable (i.e., distinguishable from finite-$N$-induced fluctuations) for bursts of $|\vrho|$, and approximately $0$ for steady incoherent states without any such burst. Hence (d) indicates that, by the end of the simulation in panel (a), due to IMR the critical coupling strength has been reset to $K_c\approx 1.75$. See text for more details.
}
\label{fig:K1.4}
\end{figure*} 

\subsection{Macroscopic bursts of coherence}\label{sec:4TTT}

To examine the aforementioned question, we first consider the following setup: We initialize the system to a minimally stable incoherent equilibrium distribution by setting $C_{min}=0$ for each agent (see Sec.\ref{sec:3}). This initial setup will be invariant to evolution with a coupling strength of $K<\Kcm$. We then consider a sudden increase in $K$ to a value satisfying $\Kcm<K<\Kcp$. 

The dynamics observed following this change of $K$ is represented in Fig. \ref{fig:K1.4}(a), for a numerical simulation of $N=10^6$ agents in $D=4$ dimensions, initialized to the minimally stable distribution with $C_{min} =0$, and then numerically integrated according to Eq. (\ref{eq:Ddimkuramoto}) with a coupling strength of $K=1.4>\Kcm\approx 1.064$ (see Table \ref{tab:Kcs}). In Fig. \ref{fig:K1.4}(a) we plot two quantities --- in the orange solid curve we present $|\vrho(t)|$, and in the blue dashed curve we show the average value of $C_{min}$ over all agents, $\langle C_{min} \rangle$. Note the rapid rise and fall of $|\vrho|$ which is accompanied by a change in value of $\langle C_{min} \rangle$. This rapid change in $\langle C_{min} \rangle$ indicates the evolution of the system away from the initialized incoherent distribution (constructed to have $\langle C_{min}\rangle \approx 0$) to a different different incoherent distribution with a larger value of $\langle C_{min}\rangle$. 

To explain the origin and consequences of this short-lived macroscopic burst of $|\vrho|$ we describe the evolution of the system in the space of distribution functions.
Let us consider a given incoherent steady state, corresponding to a distribution $F$ and having a corresponding critical coupling stability strength $K_c$ for $\Kcm\leq K_c < \Kcp$ (in the numerical example presented above, $F$ was constructed to be a minimally stable distribution with $K_c=\Kcm$).
Denote a distribution of agents for a system initialized close to this incoherent steady state by $F+\delta F$, for some perturbation $\delta F$. We then examine the expected dynamics for evolution of the system under the dynamics of Eq. (\ref{eq:Ddimkuramoto}) for a coupling strength $K$ abruptly increased from $K<K_c$ to $K_c<K<\Kcp$.

For almost every perturbation $\delta F$, the distribution $F+\delta F$ will no longer lie in the manifold of incoherent states $\scM$.
Since the initially chosen incoherent state is unstable at the increased value of $K$, for small $t$ the system will rapidly evolve away from the initial distribution, $F+\delta F$, at a rate governed by Eq. (\ref{eq:scalardispersion}), with the perturbation $\delta F$ increasing as $\delta F e^{st}$, $\Re(s)>0$. This corresponds to increasing distance away from the manifold of incoherent states, $\scM$, and hence appears as the sharp increase in $|\vrho|$ described earlier (orange curve in Fig. \ref{fig:K1.4}(a)).
Note, however, that for $K\leq \Kcp$ the analysis in Ref.\onlinecite{Chandra2018} shows that are no time-independent attractors with $|\vrho|>0$, and, further, our numerical experiments indicate that there are no $|\vrho|>0$ time-dependent attractors (e.g., periodic or chaotic).
Hence the distribution function must evolve to a stable steady-state distribution function on the manifold $\scM$. Thus, in the space of distribution functions, the evolution of the system will follow a trajectory that begins near the initial incoherent steady state in $\scM$, moves away from $\scM$, and is then attracted back towards $\scM$, but to a different incoherent steady state (corresponding to some distribution $F_1$) that is stable for the chosen coupling strength $K$. Thus, observing this system at large finite $N$ via the order parameter demonstrates an initially small value of $|\vrho|$ near zero, which rapidly rises to a large (macroscopic) value, and then falls back to a small value near zero as depicted in the representative illustration Fig. \ref{fig:K1.4}(a). 

This transition from the distribution $F\in\scM$ to the distribution $F_1\in\scM$ with $F_1\neq F$ is not distinguishable solely by observation of the time-asymptotic values of $\vrho$, since both distributions correspond to incoherent steady states. However, a signature of this transition is displayed in the transient dynamics of the macroscopic observable $\vrho$ in the form of a rapid short-lived burst of $|\vrho|$ away from its steady state value near zero.

\subsection{Instability-Mediated Resetting}\label{sec:4IMR}

An important expected consequence of the above described behavior is an `Instability-Mediated Resetting' of the system stability properties, which we define and describe as follows: The critical coupling constant of $F_1$, denoted $K_c^{(1)}$, is necessarily greater than $K$. Hence, due to the evolution of the system from $F\in\scM$ to $F_1\in\scM$ the critical coupling strength of the system has been reset from $K_c < K$ to $K_c^{(1)} > K$. This change in critical coupling strength without change in the time-asymptotic macroscopic steady-state of the system (i.e., the system is on the manifold $\scM$ corresponding to $|\vrho|=0$ at the initial state and at the asymptotic final state) is what we call \emph{Instability-Mediated Resetting}. To demonstrate this change in critical coupling strength we choose the resulting distribution at the end of the aforementioned simulation (corresponding to time $t=500$ in Fig. \ref{fig:K1.4}(a)) as the initial distribution for the following two situations: (i) evolution with $K=1.6<\Kcp\approx 2.128$, corresponding to Fig. \ref{fig:K1.4}(b), and (ii) evolution with $K=2.0<\Kcp$, corresponding to Fig. \ref{fig:K1.4}(c). Note that in Fig. \ref{fig:K1.4}(b) $|\vrho|$ and $\langle C_{min}\rangle$ do not change significantly, whereas in Fig. \ref{fig:K1.4}(c) for $K=2.0$ we see a characteristic short burst of $|\vrho(t)|$, accompanied by a change in $\langle C_{min}\rangle$. Thus we infer that $1.6<K_c^{(1)}<2.0$, hence indicating this instability-mediated resetting of the critical coupling constant for instability. To more precisely pin down the value of $K_c^{(1)}$, we evolve the system for a range of values of $K$ with each evolution having the initial condition described earlier. In Fig. \ref{fig:K1.4}(d) we plot the maximum value of $|\vrho|$ attained during the evolution as a function of $K$. We interpret Fig. \ref{fig:K1.4}(d) as follows: For all values of $K<K_c^{(1)}$ there is no burst in $|\vrho|$ and hence the maximum value is near zero; for $K>K_c^{(1)}$ the burst in $|\vrho|$ results in a large value of this maximum, and this transition from zero indicates a value of $K_c^{(1)}\approx 1.75$.

\begin{figure}
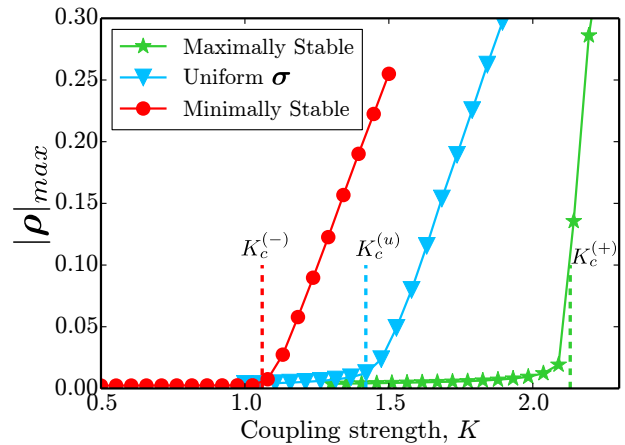

\includegraphics[width=\columnwidth]{{{Bump_transitions_m_u_p}}}
\caption{Transitions demonstrating the results derived in Eq. (\ref{eq:Kcsmup}) for a system with $N=10^6$. For the minimally stable distribution (red circles), the uniform $\vx$ distribution (blue triangles), and the maximally stable distribution (green stars), the system is evolved for various values of $K$. The maximum value attained by $|\vrho(t)|$ over a short evolution is shown as a function of $K$. For incoherent steady states that undergo stable evolution at a given value of $K$, $|\vrho|_{max}$ is approximately zero, whereas instability of incoherent steady states results in a short-lived burst of coherence, resulting in a larger value of $|\vrho|_{max}$. The theoretical predictions for the transitions to instability are shown in the respective colors using vertical dashed lines, and agree well with the numerical results. (Note that for $K>\Kcp$, $|\vrho|_{max}$ corresponds to the stable state of $|\vrho|>0$ shown in Fig. \ref{fig:rhovK} as opposed to the peak value during these short bursts.) }
\label{fig:Kcsmup}
\end{figure}

We use a similar setup to verify Eq. (\ref{eq:Kcsmup}). We consider three series of numerical simulations, corresponding to initial conditions of the minimally stable distribution (constructed with $\langle C_{min}\rangle=0$), the uniform $\vx$ distribution (setup as described in Sec. \ref{sec:3}), and the maximally stable distribution (constructed with $\langle C_{min}\rangle=1$). For each initial condition, we evolve the system with an abrupt increase from a coupling strength less than $0.5$ at $t=0$ to a given value of $K$ and note the maximum value of $|\vrho|$ attained during the evolution $t\geq 0$. This is then repeated for the same initial condition with a different value of $K$, over a range of values for $K$. The results are then plotted for this maximum attained value of $|\vrho(t)|$ as a function of $K$. As earlier, for $K$ below the corresponding $K_c$ this maximum value will be approximately zero, and for $K$ above $K_c$ the rapid macroscopic burst of $|\vrho|$ will be apparent with a larger maximum value of $|\vrho|$. Thus we expect the onset of such transient bursts for the three cases at the theoretically described values $\Kcm$, $\Kcu$ and $\Kcp$, respectively, according to Eq. (\ref{eq:Kcsmup}). These values have been marked with the vertical dashed lines in Fig. \ref{fig:Kcsmup}. Note the close agreement between these theoretically predicted values and the numerically observed burst onset. We expect improving agreement with increasing $N$. (Note that for $K>\Kcp$ the maximum attained value corresponds to the stable state of $|\vrho|>0$ as opposed to the rapid rise and fall described earlier.)

\begin{figure}
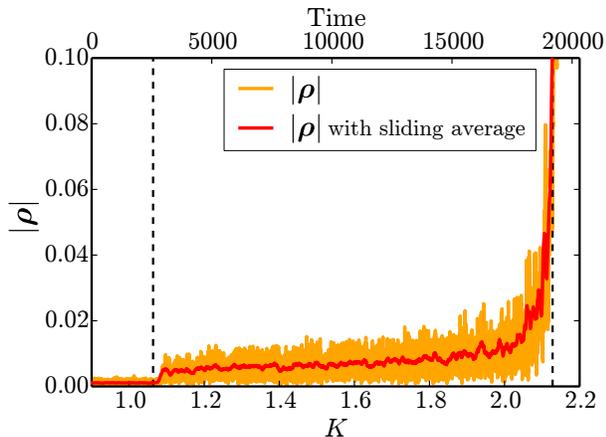

\includegraphics[width=\columnwidth]{{{IMR_var_K_minimal_N1000000_T25000_with_smooth}}}
\caption{Evolution of $|\vrho(t)|$ for a system having $N=10^6$ initialized at a minimally stable incoherent steady state with slow temporally linear increase in $K$ shown in orange, and a sliding average shown in red. The temporal increase of $K$ is linear in time and is indicated by the horizontal axis at the top of the figure panel. The vertical dashed lines correspond to $\Kcm$ and $\Kcp$. For $K\leq\Kcm$ the initialized steady state is stable and hence $|\vrho|$ maintains a value close to zero. For $\Kcm <K<\Kcp$ the system demonstrates enhanced fluctuations of $|\vrho|$ about increased, nonzero values that are apparently sustained by the continuous increase of $K$. For $K\geq \Kcp$ no incoherent state is stable, and $|\vrho|$ attains a larger value similar to Fig. \ref{fig:rhovK}}.
\label{fig:increasingK}
\end{figure}

In each of the above cases, for a system initialized to a distribution $F$, with a corresponding critical instability coupling strength of $K_c$, we examined the case of an abrupt increase in $K$ from a value of $K<K_c$ to a value $K>K_c$. The distribution $F$ remains invariant to evolution for $K<K_c$, and then, after the abrupt increase, there is an initial repulsion away from the state with distribution $F$, followed by an attraction back towards an invariant state with distribution $F_1\in\scM$. 

While the system state is away from $\scM$ and is being attracted towards $F_1$, if the value of $K$ is altered again to one greater than $K_c^{(1)}$, then the system will again be repelled away from $\scM$. As the system is then attracted towards another distinct distribution $F_2$ (with a critical coupling strength of $K_c^{(2)} > K_c^{(1)}$), the coupling strength can be varied again to $K>K_c^{(2)}$, resulting in additional delay in the attraction towards $\scM$. In this fashion, if we consider a slowly increasing coupling strength, then we can delay this attraction towards the manifold $\scM$ for large amounts of time, resulting in $|\vrho(t)|>0$ for extended periods of time without any such steady state existing at the corresponding coupling strength. This phenomenon is demonstrated in Fig \ref{fig:increasingK}, where we consider a linearly increasing coupling constant $K$, and plot $|\vrho|$ as a function of $K$ and time. We observe $|\vrho|$ to show a small fluctuating increase at $K\approx \Kcm$, which is sustained until $K\approx \Kcp$, after which $|\vrho|$ approaches the steady state value of $|\vrho|>0$ shown in Fig. \ref{fig:increasingK}. If we consider successively slower rates of increase of $K$, the resulting plot of $|\vrho|$ as a function of $K$ displays smaller fluctuations from $|\vrho|=0$ sustained through $\Kcm<K<\Kcp$. In the limit of an infinitely slow rate of increase of $K$, we expect that $|\vrho|$ will remain at zero for all $K<\Kcp$, reproducing Fig. \ref{fig:rhovK}.

\subsection{Resetting due to finite size}\label{sec:4finite}

\begin{figure*}
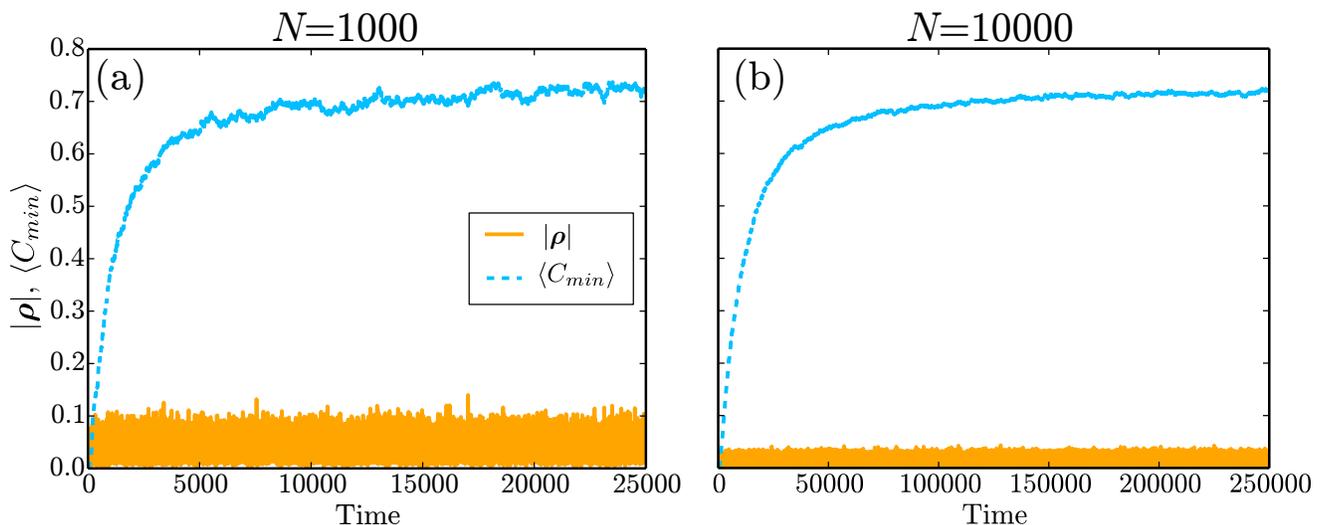

\includegraphics[width=2.\columnwidth]{{{IMR_slow_evolution_both_N}}}
\caption{Slow finite-$N$-induced evolution of the incoherent steady states for (a) $N=10^3$ and (b)$N=10^4$. Note the significantly longer timescales shown here as compared with the Fig. \ref{fig:K1.4}. $|\vrho|$ is shown as the orange curve, and $\langle C_{min}\rangle$ is shown as the blue dashed curve. Further, note the larger timescale for (b) as compared with (a). In both cases the system was initialized to a minimally stable incoherent steady state distribution with $\langle C_{min}\rangle=0$. $|\vrho|$ remains approximately zero, indicating that the system remains in $\scM$, but the gradual increase in $\langle C_{min}\rangle$ indicates the change in state on $\scM$. The final distribution achieved after long time evolution is a maximally stable state. 
}
\label{fig:finiteN}
\end{figure*} 

So far we have restricted our analysis to the $N\to\infty$ limit, wherein several incoherent equilibria in the manifold $\scM$ can simultaneously be stable to perturbations orthogonal to $\scM$ and are neutrally stable to perturbations in $\scM$. Hence, a small perturbation within $\scM$ can move an incoherent equilibrium in $\scM$ to another nearby incoherent equilibrium in $\scM$, and many such small perturbations can cumulatively cause a large change from an initial incoherent state. As we have observed earlier, transient dynamics away from $\scM$ appear to shift the critical coupling strength for loss of stability towards $K_c^{(+)}$. Thus, we suspect that perturbations away from $\scM$ are biased towards maximally stable states.
Since, in practice, $N$ is always finite it is of interest to consider the effect of finite $N$. Viewing the difference between $N$ finite but large and $N\to\infty$ as small, we can regard the system with $N$ finite but large as being akin to the $N\to\infty$ limit system with small added perturbations. Thus we might suspect finite, large $N$ to induce a slow secular evolution of the $N\to\infty$ incoherent equilibria towards a maximally stable state within $\scM$. 
In particular, we observe that for large-but-finite $N$, a system initialized at any stable incoherent equilibrium undergoes slow evolution to a equilibrium corresponding to a maximally stable state. We demonstrate this effect in Fig. \ref{fig:finiteN}(a), where we plot $\langle C_{min} \rangle$ as a function of time for evolution of $N=10^3$ agents initialized to the minimally stable distribution with $\langle C_{min}\rangle=0$, evolved with $K=0.9<\Kcm$. Note that  $\langle C_{min} \rangle$ undergoes slow growth and eventually asymptotes to a large value of  $\langle C_{min} \rangle\approx 0.7$ at very long times. After this long time, the large value of  $\langle C_{min} \rangle$ indicates that a large fraction of agents have moved to the subspace corresponding to the lowest frequency of rotation, similar to our setup of the maximally stable distribution in Sec. \ref{sec:3}. From this state if we consider sudden changes in $K$ to values in the range of $\Kcm<K<\Kcp$, we do not observe any characteristic burst in the value of $|\vrho|$, indicating that our distribution was indeed a maximally stable distribution.

Since this evolution towards a maximally stable distribution appears to be mediated by finite-$N$ effects, we expect this evolution to become progressively slower as $N$ increases, with stationarity of the incoherent states restored as $N\to\infty$. This picture is confirmed numerically Fig. \ref{fig:finiteN}, where it can be clearly seen that for a larger value of $N=10^4$ initialized as earlier with  $\langle C_{min} \rangle=0$ and evolved at $K=0.9<\Kcm$ takes about ten times longer time to reach an asymptotic state for  $\langle C_{min} \rangle$ (Note the different scales on the x-axes of the plots). 
Thus, for $N$ large but finite, if one were to initialize an incoherent equilibrium state with $K<K_c$ (where $K_c$ is calculated in the $N\to\infty$ limit) and wait for sufficiently long time, then one could continuously increase $K$ without the incoherent state becoming unstable until $K$ reaches $K_c^{(+)}$.

\section{Conclusions}\label{sec:5}

In this paper we look at a $D$-dimensional generalization of the Kuramoto model\cite{Chandra2018}. Unlike the case of the standard ($D=2$) Kuramoto model, we have shown that for even $D\geq 4$ there are an infinite number of time-independent distributions of agents (defining the manifold $\scM$) that correspond to the completely incoherent state (i.e., having $|\vrho|=0$) in the infinite system-size limit (Sec. \ref{sec:2}). We then proved that these distributions demonstrate different stabilities, with each distribution being stable for coupling strengths below a critical coupling strength $K_c$ corresponding to the given distribution (Sec. \ref{sec:3}). Further, for each value of $K_c$ within a range $\Kcm<K_c<\Kcp=2\Kcm$ there exists an infinite number of distributions that become unstable as $K$ is increased through $K_c$. In Sec. \ref{sec:4} we show that these properties result in transitions within the $|\vrho|=0$ manifold $\scM$ of steady states as $K$ is increased in the range $[\Kcm,\Kcp]$, which leave their signatures as short-lived macroscopic bursts in the value of $|\vrho|$ (Fig. \ref{fig:K1.4}). These transitions imply a change in the microscopic state of the system, with the system state after a transient having a significantly larger critical coupling strength for instability due to an Instability-Mediated Resetting of the distribution function (Fig. \ref{fig:K1.4}(d)). While for all $K<\Kcp$ the only stable steady states are on $\scM$, we demonstrate (Fig. \ref{fig:increasingK}) that considering a linearly increasing $K$ results in a small positive fluctuating value of $|\vrho(t)|$ (and hence indicating evolution not on $\scM$) which can be sustained for long periods of time as $K$ is linearly increased through the range $K_c<K<\Kcp$ (where $K_c$ refers to the originally initialized distribution); also, these fluctuations in $|\vrho|$ become smaller as the rate of increase of $K$ with time becomes slower.
Since there are a multitude of stable states on $\scM$, with neutral stability to perturbations in $\scM$, noise can cause slow evolution of states in $\scM$. We observe such slow evolution due to noise induced by finite-$N$ effects (Fig. \ref{fig:finiteN}), which evolves the system towards a maximally-stable distribution.

\section{Acknowledgement}
We thank Michelle Girvan and Thomas M. Antonsen for useful discussion. This work was supported by ONR grant N000141512134 and by AFOSR grant FA9550-15-1-0171

\bibliography{IMRbib}
\end{document}